\documentclass{emulateapj}
\usepackage{apjfonts}

\newcommand{\Msun}      {\mbox{$\rm\,M_{\mathord\odot}$}}

\begin{document}

\lefthead{4U 1730--22: A Neutron Star in Quiescence}
\righthead{Tomsick, Gelino, \& Kaaret}

\submitted{Accepted by the Astrophysical Journal}

\title{Uncovering the Nature of the X-ray Transient 4U~1730--22: Discovery 
of X-ray Emission from a Neutron Star in Quiescence with {\em Chandra}}

\author{John A. Tomsick\altaffilmark{1,2},
Dawn M. Gelino\altaffilmark{3},
Philip Kaaret\altaffilmark{4}}

\altaffiltext{1}{Space Sciences Laboratory, 7 Gauss Way, University
of California, Berkeley, CA 94720-7450
(e-mail: jtomsick@ssl.berkeley.edu)}

\altaffiltext{2}{Center for Astrophysics and Space Sciences, 
9500 Gilman Drive, Code 0424, University of California at San Diego, 
La Jolla, CA 92093}

\altaffiltext{3}{Michelson Science Center, California Institute of 
Technology, 770 South Wilson Avenue, MS 100-22, Pasadena, CA 91125}

\altaffiltext{4}{Department of Physics and Astronomy, University of
Iowa, Iowa City, IA 52242}

\begin{abstract}

The X-ray transient, 4U~1730--22, has not been detected in outburst 
since 1972, when a single $\sim$200 day outburst was detected by the
{\em Uhuru} satellite.  This neutron star or black hole X-ray binary
is presumably in quiescence now, and here, we report on X-ray and
optical observations of the 4U~1730--22 field designed to identify
the system's quiescent counterpart.  Using the {\em Chandra X-ray 
Observatory}, we have found a very likely counterpart.  The candidate 
counterpart, CXOU J173357.5--220156, is close to the center of the 
{\em Uhuru} error region and has a thermal spectrum.  The 0.3--8 keV 
spectrum is well-described by a neutron star atmosphere model with an 
effective temperature of $131\pm 21$ eV.  For a neutron star with a 
10 km radius, the implied source distance is $10^{+12}_{-4}$ kpc, and 
the X-ray luminosity is $1.9\times 10^{33}$ ($d$/10 kpc)$^{2}$ ergs~s$^{-1}$.  
Accretion from a companion star is likely required to maintain the 
temperature of this neutron star, which would imply that it is an X-ray 
binary and therefore, almost certainly the 4U~1730--22 counterpart.  
We do not detect an optical source at the position of the {\em Chandra}
source down to $R > 22.1$, and this is consistent with the system
being a Low-Mass X-ray Binary at a distance greater than a few kpc.
If our identification is correct, 4U~1730--22 is one of the 5 most
luminous of the 20 neutron star transients that have quiescent X-ray
luminosity measurements.  We discuss the results in the context of
neutron star cooling and the comparison between neutron stars and
black holes in quiescence.

\end{abstract}

\keywords{accretion, accretion disks --- stars: neutron ---
stars: individual (4U~1730--22) --- X-rays: stars --- X-rays: general}

\section{Introduction}

{\em Uhuru}, the first X-ray astronomy satellite \citep{giacconi71}, was
launched in 1970 and carried out an all-sky survey \citep{forman78}.  
Although the 339 X-ray sources it detected are bright by today's standards,
and most of them have been well-studied, some of the transient sources 
have not been studied by subsequent X-ray satellites.  Here, we focus on 
one such transient, 4U~1730--22, that was discovered by {\em Uhuru} in 
1972 when it entered into an X-ray outburst lasting $\sim$200 days 
\citep{cominsky78,csl97}.  The outburst behavior and X-ray spectrum 
measured by {\em Uhuru} showed all of the signs that the source is an 
X-ray binary.  \cite{ts96} and \cite{csl97} both classify this system 
as a probable neutron star system, presumably based on the {\em Uhuru} 
X-ray spectrum, but there is no information from type I X-ray bursts, 
pulsations, or a compact object mass measurement to indicate whether 
the system contains a neutron star or a black hole.  

In quiescence, neutron star and black hole soft X-ray transients (SXTs), 
which are normally Low-Mass X-ray Binaries (LMXBs), typically have X-ray 
luminosities in the $10^{30}$ to $10^{34}$ ergs~s$^{-1}$ range.  Although 
these are faint sources, observations with the {\em Chandra X-ray Observatory} 
are able to probe much of this range for Galactic sources at distances
out to $\sim$10--15 kpc.  Here, we report on the results of {\em Chandra}
and optical observations covering the {\em Uhuru} error region for
4U~1730--22.  One reason for choosing this source is that its location
($l = 4.5^{\circ}$, $b = +5.9^{\circ}$) and its outburst X-ray spectrum
suggest that it is not highly absorbed, making it more likely to be
detected in the X-ray band and easier to study in the X-ray and optical.

The purpose of this work is both to learn more about 4U~1730--22 as well
as about neutron stars and black holes in quiescence.  Observations of
these sources in quiescence provide some of the best opportunities to
probe the properties of the compact objects themselves.  For most 
neutron star transients, a thermal component with a temperature of
$\sim$0.1 keV is present in the X-ray spectrum, and it is thought that
this emission is coming directly from the surface of the neutron star 
\citep{campana98,bbr98,rutledge99}.  This provides an opportunity to
study the short time scale (years) cooling of the neutron star crust
\citep{rutledge02,wijnands05} as well as probing the core temperature, 
which changes on a time scale of millennia \citep{colpi01}.  If 
4U~1730--22 harbors a neutron star, we expect that the core temperature
will set its X-ray emission properties as it has not had an outburst 
(as far as we know) for 34 years.  For black holes, the lack of a thermal 
component as well as the fact that, on average, quiescent black holes tend 
to be fainter than neutron stars have been taken as possible evidence
for the existence of black hole event horizons \citep{ngm97,garcia01,mnr04}.
Recent observations of some very faint neutron star systems 
\citep{tomsick04_2123,tomsick05,jonker06} make it important to obtain
X-ray measurements for more quiescent SXTs.

\section{Analysis and Results}

\subsection{{\em Chandra} Reduction and Source Detection}

\begin{figure}
\centerline{\includegraphics[width=0.50\textwidth]{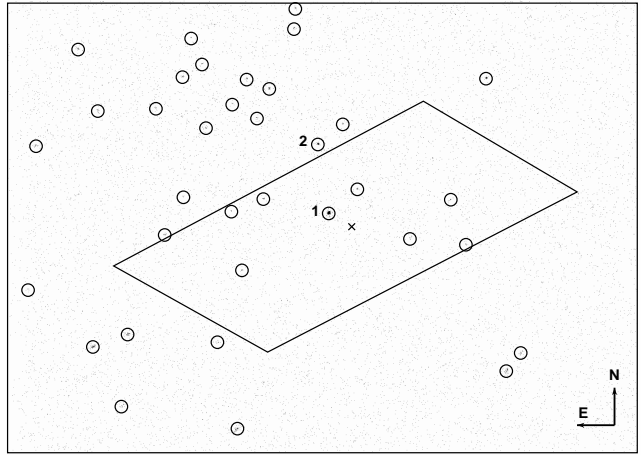}}
\caption{The 0.3--8 keV ACIS image of the 4U~1730--22 field from a 39.8 ks
{\em Chandra} observation.  For scale, the N and E arrows are $30^{\prime\prime}$
in length, and the full $6^{\prime}$-by-$8^{\prime}$ region searched for sources
is shown.  The parallelogram shows the 90\% confidence {\em Uhuru} error region 
for 4U~1730--22, and the cross is the best {\em Uhuru} position.  The 35 detected
{\em Chandra} sources are marked with circles, and sources 1 and 2 (described
in the text) are labeled.\label{fig:chandra_image}}
\end{figure}

We observed the 4U~1730--22 field in the X-ray band with {\em Chandra} on 
2004 May 11 (ObsID 4583), obtaining an exposure time of 39.8 ks.  We used 
the Advanced CCD Imaging Spectrometer \citep[ACIS,][]{garmire03} with the 
target position placed on the ACIS-S3 chip.  To obtain the lowest possible
background, the observation was performed with the ACIS in ``VFAINT" readout 
mode, which provides the maximum amount of information per event.  To 
prepare the data for our analysis, we started with the ``level 1" event 
list, which was produced using standard processing with ASCDS version 
7.6.7.2.  We further processed the data using the {\em Chandra} Interactive 
Analysis of Observations (CIAO) version 3.3.0.1 software and the Calibration 
Data Base (CALDB) version 3.2.3.  A portion of the image from the resulting 
``level 2" event list is shown in Figure~\ref{fig:chandra_image}.  

After inspecting the full ACIS field-of-view (six $8^{\prime}$-by-$8^{\prime}$
chips), we used the CIAO routine {\ttfamily wavdetect} \citep{freeman02} to 
search for potential 4U~1730--22 counterparts in or near the {\em Uhuru} 
position for this source.  For the search, we restricted the event list 
energy range to 0.3--8 keV and the region to $6^{\prime}$ in the North-South 
direction and $8^{\prime}$ in the East-West direction.  The search region is 
centered on the best {\em Uhuru} position for 4U~1730--22 \citep{forman78} 
and includes a region considerably larger than the 90\% confidence error 
region for 4U~1730--22, which is shown in Figure~\ref{fig:chandra_image}.  
When running {\ttfamily wavdetect}, we used a detection threshold of 
$1.4\times 10^{-6}$ so that we expect that $\sim$1 source that we 
detect in the $8^{\prime}$-by-$6^{\prime}$ region will be spurious.  

Upon running {\ttfamily wavdetect}, we find 35 sources in the full
$8^{\prime}$-by-$6^{\prime}$ region, and these are marked with circles 
in Figure~\ref{fig:chandra_image}.  In Table~\ref{tab:sources}, the 
sources are listed along with their positions in order of their brightness 
(the number of counts per source estimated by {\ttfamily wavdetect}.
The brightest source detected (source 1, CXOU J173357.5--220156) has 
545 counts and is also the closest source to the best {\em Uhuru} position, 
being only $0^{\prime}\!.36$ away.  The next brightest source (source 2,
CXOU J173358.1--220101) is considerably fainter with 88 counts, and it is 
$1^{\prime}\!.17$ from the best {\em Uhuru} position.  Of the 35 sources, 
8 of them lie within the 90\% confidence {\em Uhuru} error region and 5 
more lie within $30^{\prime\prime}$ of the error region.  Thus, there are 
13 sources that should be considered as possible 4U~1730--22 counterparts 
based on their positions.  However, we will only be able to obtain further 
X-ray information for sources 1 and 2 as we only detected between 4 and 12 
counts per source for the other 11 sources.  The following sub-section 
describes the X-ray spectral and timing analysis for the two brightest 
sources.

\subsection{{\em Chandra} Spectral and Timing Results for the Two Brightest Sources}

We extracted spectra for the two brightest sources in the 4U~1730--22 field 
described above.  In both cases, we extracted counts from a $5^{\prime\prime}$ 
radius circle centered on the source position determined by {\ttfamily wavdetect} 
(see Table~\ref{tab:sources}).  We also extracted background spectra from an 
annulus centered on the source position with inner radius of $10^{\prime\prime}$ 
and an outer radius of $50^{\prime\prime}$.  We used the CIAO routine 
{\ttfamily psextract} to produce the spectra as well as the response matrices.  
For the two sources, we collected 566 and 95 counts (in the 0.3--8 keV band), 
respectively, and, in each case, we estimate background levels of between 6 and 
7 counts.  We binned the spectra so that the source 1 spectrum has an average
of 31 counts per bin and the source 2 spectrum has an average of 12 counts per
bin.

\begin{figure}
\centerline{\includegraphics[width=0.55\textwidth]{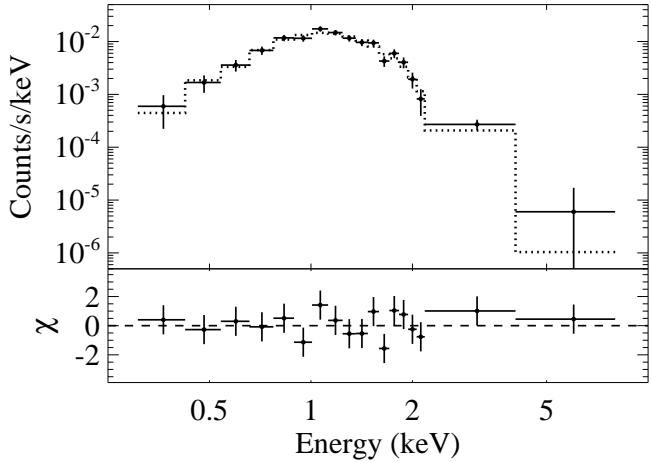}}
\vspace{-1.3cm}
\caption{The {\em Chandra}/ACIS spectrum for source 1.  The top panel shows the 
counts spectrum fitted with a neutron star atmosphere (NSA) model (dotted line), 
and the bottom panel shows the residuals for this fit.\label{fig:spec1}}
\end{figure}

\begin{figure}
\centerline{\includegraphics[width=0.55\textwidth]{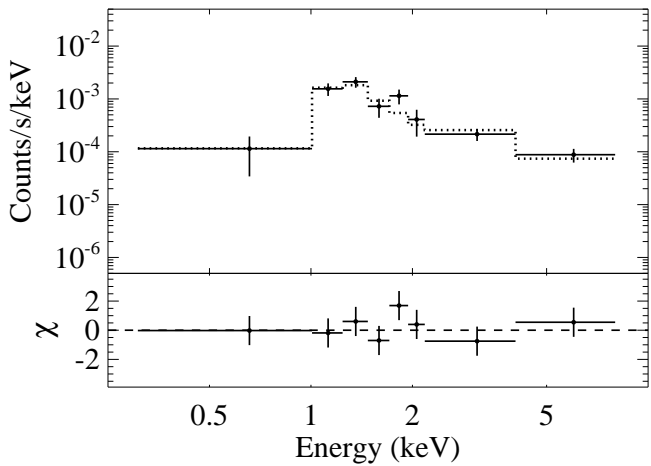}}
\vspace{-1.3cm}
\caption{{\em Chandra}/ACIS spectrum for source 2.  The top panel shows the counts
spectrum fitted with a two-component NSA plus power-law model (dotted line), and 
the bottom panel shows the residuals for this fit.\label{fig:spec2}}
\end{figure}

The energy spectra for the two sources are shown in Figures~\ref{fig:spec1}
and \ref{fig:spec2}, and we used the XSPEC version 11.3.2o software package
to perform model fitting.  We began by using single component models, including 
power-law, blackbody, and bremsstrahlung models.  In each case, we accounted
for interstellar absorption, using the photo-electric absorption cross sections 
from \cite{bm92} and elemental abundances from \cite{ag89}, and we left the 
column density ($N_{\rm H}$) as a free parameter.  For source 1, we obtained 
good fits for all 3 models with values of $\chi^{2}$ between 10 and 13 for 
15 degrees of freedom.  The power-law necessary to fit the spectrum is extremely 
steep with a photon index of $\Gamma = 5.4^{+0.6}_{-0.5}$ (all errors are 90\%
confidence unless specified otherwise), and this likely indicates that the 
emission has a thermal origin.  The blackbody and bremsstrahlung temperatures 
are $0.27\pm 0.02$ and $0.52^{+0.09}_{-0.08}$ keV, respectively.  Of the 3 models, 
only the blackbody model results in a column density 
$N_{\rm H} = (2.6\pm 0.7)\times 10^{21}$ cm$^{-2}$ that is consistent with the 
value of $3\times 10^{21}$ cm$^{-2}$ measured along this line-of-sight through 
the Galaxy \citep{dl90}.  We obtain column densities of 
$(8.6^{+1.4}_{-1.2})\times 10^{21}$ and $(4.9^{+0.9}_{-0.7})\times 10^{21}$ 
cm$^{-2}$ for the power-law and bremsstrahlung models, respectively.

Using the blackbody model fit to the source 1 spectrum, we estimate that
the 0.3--8 keV absorbed flux is $4.9\times 10^{-14}$ ergs~cm$^{-2}$~s$^{-1}$, 
while the unabsorbed flux is $9.8\times 10^{-14}$ ergs~cm$^{-2}$~s$^{-1}$.
The $\log{N}$-$\log{S}$ curve from \cite{virani06} allows us to estimate how
likely it is to find a source as bright as source 1 by chance.  As 
\cite{virani06} compile the $\log{N}$-$\log{S}$ using {\em Chandra} data from
an observation well out of the Galactic plane ($b\sim -54^{\circ}$), 
their source population is dominated by Active Galactic Nuclei (AGN).  
In addition, the high Galactic latitude used for their study means that
it is appropriate for us to consider the unabsorbed flux for source 1, 
which is $8.1\times 10^{-14}$ ergs~cm$^{-2}$~s$^{-1}$ in the 0.5--2 keV 
energy band.  At this flux level, the AGN source density is near 
5 sources per square degree.  Considering the 9.4 square arcminute size
of the 4U~1730--22 {\em Uhuru} error circle, approximately 0.01 AGN
as bright as source 1 are expected to be detected by chance.  In addition
to the low probability of finding an AGN at this flux level, the spectral
shape of source 1 would be extremely ususual for an AGN \citep{tozzi06},
and we conclude that it is very unlikely that source 1 is an AGN.  In 
addition, it is unlikely that source 1 is due to coronal emission from a 
nearby, active star.  Although the bremsstrahlung temperature derived from 
the spectral fits is consistent with such a hypothesis, one would also expect 
to see strong emission lines \citep{canizares00}, which are not present.

The steep spectrum ($\Gamma\sim 5$) described above for source 1 could be 
produced by a quiescent neutron star system, but it would be unprecedented 
for a black hole transient in quiescence \citep{kong02a,ctk06}.  Thus, 
we refitted the spectrum with a neutron star atmosphere (NSA) model, which 
is the spectrum for thermal emission from the hydrogen atmosphere of a neutron 
star \citep{psz91,zps96}.  For the NSA fit, we assume a weakly magnetized 
($B < 10^{8} - 10^{9}$ G) neutron star with a canonical radius and mass of 
10~km and 1.4\Msun, respectively.  Fitting the spectrum yields an effective 
temperature of $(1.51\pm 0.24)\times 10^{6}$ K, corresponding to 
$kT_{\rm eff} = 131\pm 21$~eV.  Such a temperature is in-line with the range 
of values that have been seen for neutron star soft X-ray SXTs in quiescence 
\citep{rutledge99,wijnands04,tomsick04_2123}.  The normalization for the NSA 
model is directly related to the source distance ($d$), and in this case, the 
value given in Table~\ref{tab:spectra} implies $d = 10^{+12}_{-4}$~kpc 
(90\% confidence).  Although this is a large range of possible values, the 
values are consistent with what is expected for a Galactic X-ray binary.  
Finally, some quiescent neutron star systems require two-component models
\citep{rutledge99,jonker04}; however, fitting the source 1 spectrum with the 
combination of NSA and power-law models does not lead to a significant 
improvement in the quality of the fit (see Table~\ref{tab:spectra}).

Although the quality of the source 2 spectrum is significantly lower than
for source 1, we used the same approach in fitting the source 2 spectrum.
All three of the simple one-component models provide relatively poor fits
with power-law, blackbody, and bremsstrahlung models yielding $\chi^{2}$
values of 12, 20, and 14, respectively, for 5 degrees of freedom.  When 
using the power-law model, the measured photon index is $\Gamma = 
2.1^{+1.2}_{-0.9}$, indicating that source 2 is significantly harder than 
source 1.  The fact that the spectrum is quite hard explains why a 
blackbody provides an even worse fit than the power-law model.  Using
the power-law model with $N_{\rm H}$ fixed to the Galactic value, we
infer an unabsorbed 0.5--2 keV flux of $6.0\times 10^{-15}$ 
ergs~cm$^{-2}$~s$^{-1}$.  At this flux, the AGN density is close to 
100 sources per square degree \citep{virani06}, implying that one would
expect $\sim$0.3 AGN at this flux level within the {\em Uhuru} error
region.  Given the lower flux of source 2 and the fact that its spectrum
is similar to that of AGN, the X-ray information allows for the possibility
that source 2 is an AGN.

To allow for a direct comparison between sources 1 and 2, we have also
fitted the source 2 spectrum with the NSA and NSA plus power-law models,
and the results are given in Table~\ref{tab:spectra}.  As for the blackbody
model, an NSA fit to the source 2 spectrum is poor.  In addition, the lower 
limit on the source distance of $d > 183$ kpc that is implied by the NSA
fit is not consistent with a Galactic source, and the lower limit on the 
temperature of $kT > 432$ eV implies a temperature above that expected for 
a quiescent neutron star.  With the power-law plus NSA model, the quality 
of the fit improves to $\chi^{2}/\nu = 4.7/3$.  The parameters with this 
model are not well-constrained, and the error regions would allow for 
reasonable NSA parameters.  However, the best fit column density of 
$2.2\times 10^{22}$ cm$^{-2}$ is higher than we expect for the 4U~1730--22 
counterpart.  Even if the source distance is fixed to a reasonable distance 
of 10~kpc, we find $N_{\rm H} > 1\times 10^{22}$ cm$^{-2}$.  This may indicate 
that the power-law plus NSA model is not the correct model of this spectrum, 
but the quality of the spectrum precludes a definite conclusion to this 
question.

We extracted light curves for the two sources to determine if either of
them show evidence for variability.  While the statistics do not allow
for a detailed timing study, there is some probability that we would see
an eclipse for a binary system.  However, inspection of the 0.3--8 keV 
light curves with 2,000~s time bins does not show any clearly significant 
variability.  For source 1, the mean number of counts per 2000~s bin is 28 
and the minimum is 19, so there is no evidence for an eclipse.  We also
tested for variability by comparing the times that X-ray events were detected 
over the 40~ks observation to a distribution with the same number of events 
spread evenly throughout the observation and using a Kolmogorov-Smirnov (KS)
test.  With this method, we find possible evidence that source 1 is variable, 
but the detection of variability is not highly significant, with an 11\% 
chance that it is spurious.  There is no evidence for variability from
source 2.

\subsection{Optical Observations}

\begin{figure}
\centerline{\includegraphics[width=0.47\textwidth]{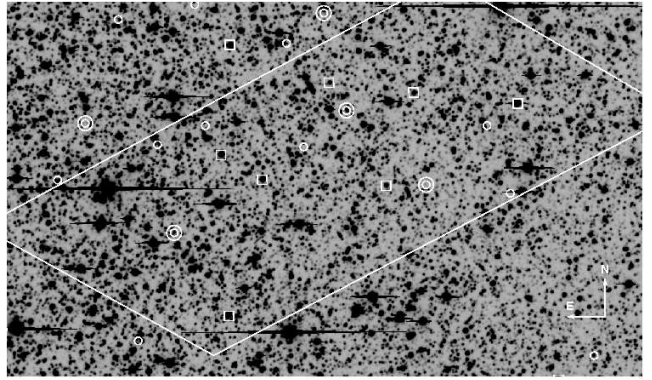}}
\caption{The $R$-band image of the 4U~1730--22 field from observations with 
the CTIO 4 meter telescope and the Mosaic II CCD imager.  The image is the
combination of 4 exposures with a total of 990 seconds of exposure time.
For scale, the N and E arrows are $20^{\prime\prime}$ in length.  The 
parallelogram is the {\em Uhuru} error region for 4U~1730--22.  The squares
mark the optical sources that we used to register the image.  The circles
mark the locations of {\em Chandra} sources, and {\em Chandra} sources with
likely optical identifications (see Table~\ref{tab:sources}) are
marked with 2 circles.\label{fig:noao_image}}
\end{figure}

We obtained optical images of the 4U~1730--22 field with the 4 meter
telescope at Cerro Tololo Inter-American Observatory (CTIO) on 2003 
March 12.  Starting at 8.14 hours UT, we obtained exposures with 
durations of 30, 60, 300, and 600 seconds with the Mosaic II CCD imager 
and the Harris $R$-band filter.  The Harris $R$-band filter is much 
more similar to a Johnson $R$-band filter than Cousins-Kron $R$-band
filter.  The conditions during the observations were relatively good, 
with sub-arcsecond seeing.  We used the Image Reduction and Analysis 
Facility (IRAF) software to reduce the images, applying bias subtraction 
and flat-fielding the images.  Due to the combination of relatively low 
extinction but proximity to the Galactic Center, the field is very 
crowded, and we had to be careful when registering the images.  To 
perform the position registration, we identified 16 stars that are in 
the USNO-B1.0 catalog with brightnesses between $R = 14$ and 18.  It 
is critical to use relatively bright stars to be sure that source 
confusion is avoided.  Using these 16 stars, we registered the images 
using IRAF and, based on comparisons between USNO positions and the 
positions we obtain after registering the images (see 
Figure~\ref{fig:noao_image}), the registration is good to 
$0^{\prime\prime}\!.4$.

For magnitude calibration, we obtained exposures of the Landolt field
PG 1323--086 \citep{landolt92}.  At 9.52 hours and 9.70 hours UT, two 
3 second $R$-band exposures were taken.  Aperture photometry with IRAF 
indicates that the count rates for these two exposures were the same to 
within 0.6\%, consistent with photometric conditions at this time.  
However, aperture photometry on several sources in the four 4U~1730--22 
exposures indicates count rate changes by $\sim$6\% between the brightest 
and faintest exposures even though the air mass only changes between 
1.20 and 1.29 during the observations.  This suggests that the 
conditions were not completely photometric during the 4U~1730--22
exposures, but the 6\% change only corresponds to a change in 
$R$-magnitude of $\sim$0.1, which is not a major concern for our 
purposes.  To calibrate the 4U~1730--22 field, we used the Landolt 
count rates and the 30 second 4U~1730--22 exposure for which the 
highest count rates were seen.  

Figure~\ref{fig:noao_image} shows the $R$-band image for all four of
our exposures combined (990 seconds) along with the positions of 
the {\em Chandra} sources in the 4U~1730--22 field.  We inspected
the $R$-band image in the regions near the 13 X-ray sources that are
in or close to the {\em Uhuru} error region.  Given that the uncertainty
in the optical position registration is $0^{\prime\prime}\!.4$, and the 
{\em Chandra} pointing uncertainty is $0^{\prime\prime}\!.6$, the optical
and X-ray positions must be within $\sim$$0^{\prime\prime}\!.7$ for a
convincing identification.  Of the 13 X-ray sources, 5 of them meet
this criterion with differences between X-ray and optical positions
between $0^{\prime\prime}\!.20$ and $0^{\prime\prime}\!.58$.  These 
sources are marked in Figure~\ref{fig:noao_image}, and their $R$-band 
magnitudes and the separations between their X-ray and optical 
positions are given in Table~\ref{tab:sources}.  All of the sources 
with optical counterparts are weak X-ray sources with between 5 and 
11 ACIS counts.  In the optical, two of the sources are relatively 
bright with $R$-band magnitudes of $16.7\pm 0.1$ and $18.9\pm 0.1$, 
and $R\sim 21.5\pm 0.5$ for the other three.

We did not detect optical counterparts for either of the bright X-ray 
sources (sources 1 and 2 above).  In fact, in both cases, the nearest 
optical source is $2^{\prime\prime}$ from the X-ray position, so that 
the nearest optical sources are clearly not associated with the X-ray 
sources.  Thus, it is important to derive the limiting magnitude for the 
$R$-band image.  All previous aperture photometry was performed using
apertures with 4 pixel ($1^{\prime\prime}\!.08$) radii; thus, we consider
an aperture with the same size.  The sky count rate is $\sim$340 counts 
per pixel, giving a background level of $\sim$17,000 counts per 
aperture.  To obtain a detection with signal-to-noise ratio of 5 
requires a source with 650 counts, and, using the magnitude calibration
described above, the limiting magnitude is $R = 22.1$.  

\section{Discussion}

\subsection{The 4U~1730--22 Counterpart}

Based on the X-ray properties of source 1 presented above, a relatively
strong case can be made that it is the 4U~1730--22 counterpart.  The
strongest evidence comes from the X-ray spectrum, which is consistent with 
a thermal spectrum (with interstellar absorption) from a standard-sized 
(10 km radius) neutron star at a distance of $\sim$10 kpc.  While it is worth
considering the possibility that the source is an isolated neutron star, 
especially given its faintness in the optical, the relatively high thermal 
temperature, $kT_{\rm eff} = 131$ eV, would indicate a very young neutron
star ($<$10,000 years) \citep{yp04}.  This is unlikely given that there
is no evidence for a supernova remnant or extended emission of any kind
in the X-ray or optical.  On the other hand, a temperature of 131 eV
is in the range of values typically seen for quiescent neutron star 
transients \citep{rutledge99,wijnands04,tomsick04_2123} as the neutron star 
temperatures in such systems are kept relatively high due to the intermittent 
accretion \citep{bbr98}.  Of course, the other reason to suspect that source 1 
is the 4U~1730--22 counterpart is that it lies very close to the center
of the {\em Uhuru} error region.  

We have also considered the possibility that source 2 might be the
4U~1730--22 counterpart.  The fact that the source is much harder than 
source 1 indicates that it does not have a purely thermal spectrum, but
this does not rule out the possibility that the source is a quiescent 
neutron star or black hole transient.  The spectra of quiescent neutron 
star transients can be purely thermal, purely non-thermal, or require
both components \citep{jonker04}.  For source 2, the problem with the
first 2 models is that they give poor fits with $\chi^{2}_{\nu} = 3.3$
for the NSA model and $\chi^{2}_{\nu} = 2.5$ for the power-law model.  
Although the power-law plus NSA model provides a better fit 
($\chi^{2}_{\nu} = 1.6$), as discussed above, the value of $N_{\rm H}$ 
is somewhat higher than one would expect for this line of sight.  
Considering the black hole possibility, it is worth noting that the 
power-law photon index we measure for source 2, $\Gamma\sim 2$, is 
typical of black hole transients in quiescence.  However, in other
{\em Chandra} observations of such systems, a power-law typically
gives a formally acceptable fit $\chi^{2}_{\nu}\sim 1$
\citep[e.g.,][]{kong02a}.  Overall, although there are some reasons
to think that source 2 is not a neutron star or black hole transient, 
a better spectrum is necessary to reach a definite conclusion.
Probably the strongest argument against source 2 being the quiescent
counterpart to 4U~1730--22 is that source 1, with its distinctive 
thermal spectrum and its position close to the center of the 
{\em Uhuru} error region, is such a good candidate.

\subsection{Possible Spectral Types and Distance Estimate}

As described above, the X-ray spectrum of source 1 is consistent with a 
quiescent neutron star with a 10 km radius at a distance of $10^{+12}_{-4}$ kpc, 
and we can check on whether the optical non-detection is consistent with the 
source being an LMXB.  Most neutron star SXTs have companions with K-type 
main sequence companions (K3V--K7V) \citep{cc06,chevalier99}.  Even when 
these sources are in quiescence, as much as half of the optical light can
come from the accretion disk \citep{tomsick02,torres02}.  Thus, in placing
limits on the source distance, we consider a range of spectral types from
M0V to K0V and disk light fractions up to $f = 0.5$.  We estimate the 
optical extinction using the measured X-ray column density of 
$N_{\rm H} = 3.7\times 10^{21}$ cm$^{-2}$, which corresponds to 
$A_{V} = 2.1$ \citep{ps95} and $A_{R} = 1.6$ \citep{ccm89}.  For a K0V star, 
the absolute optical magnitudes are $M_{V} = 5.9$ and $M_{R} = 5.3$ \citep{cox00}.  
Thus, considering that $R > 22.1$ for source 1, a K0V-type companion implies 
lower limits on the source distance of $d > 11.0$ kpc for $f = 0$ and 
$d > 15.1$ kpc for $f = 0.5$.  On the other end of the range, a M0V-type 
companion with $M_{R} = 7.7$ allows for a distance as low as $d > 3.6$ kpc 
for $f = 0$ and $d > 5.0$ kpc for $f = 0.5$.  Thus, the optical non-detection 
allows for the full range of companion spectral types we expect for a neutron 
star SXT but suggests that the source is at a distance larger than a few kpc, 
which is consistent with the lower limit of 6 kpc implied by the NSA fit to 
the X-ray spectrum.  

Given that the optical information will only provide us with lower limits
to the source 1 distance, we consider the best estimate of the distance
to be the 10 kpc from the NSA fit to the X-ray spectrum.  Such a distance
may very well be consistent with the possibility that source 1 is the
4U~1730--22 counterpart as the outburst from this source was considerably
fainter than what has been seen from other neutron star SXTs (Cen~X-4, 
4U~1608--522, and Aql~X-1).  According to \cite{csl97}, the peak 4U~1730--22 
X-ray flux during its 1972 outburst was 0.12 Crab, compared to 29.6 and 
4 Crab for 2 outbursts from Cen~X-4 and a median peak flux of 0.74 Crab 
for 24 outbursts from 4U~1608--522 and Aql~X-1.  Distance estimates for 
Cen X-4, 4U~1608--522, and Aql~X-1 are 1.2, 3.6, and 5 kpc, respectively 
\citep[][and references therein]{tomsick04_2123}.  Although this is only 
based on one outburst from 4U~1730--22, it could indicate that the source 
is a factor 2--3 times farther away than 4U~1608--522 and Aql X-1, which 
is in-line with a distance of 10 kpc.  Furthermore, this would imply a peak 
X-ray luminosity for 4U~1730--22 of $3\times 10^{37}$ ($d$/10 kpc)$^{2}$ 
ergs~s$^{-1}$, which is typical of neutron star SXTs.

\subsection{Neutron Star SXTs in Quiescence}

It is also interesting to look at how our result for source 1, which
we will henceforth call 4U~1730--22 as it appears that the association
is quite likely, fits into the larger body of work on quiescent neutron
star and black hole SXTs.  One important question concerns the significance
of the difference between neutron star and black hole luminosities.  
Early observations based on a handful of sources suggested that black 
holes were much less luminous \citep{ngm97,garcia01}, but recent
observations have shown that there is significant overlap between the
two distributions \citep{tomsick05,jonker06}.  Based on the NSA fit
to the 4U~1730--22 spectrum, we find an unabsorbed 0.3--8 keV flux of
$1.6\times 10^{-13}$ ergs~cm$^{-2}$~s$^{-1}$, which corresponds to
a luminosity of $L = 1.9\times 10^{33}$ ($d$/10 kpc)$^{2}$ ergs~s$^{-1}$.
Assuming a distance of 10 kpc and a neutron star mass of 1.4\Msun, this
corresponds to an Eddington-scaled luminosity of $L/L_{\rm Edd} = 
1.0\times 10^{-5}$, which is a factor of $\sim$6 higher than the 
median value for the 19 neutron star systems shown in Figure~5 of
\cite{tomsick05}.  Out of the group of 20 neutron star systems, there
is only one (EXO~0748--676) with a higher luminosity, and 3 others
(4U~1608--522, EXO~1745--248, and XTE~J1709--267) with comparable
luminosities, so that 4U~1730--22 has one of the 5 highest quiescent 
X-ray luminosities (assuming a distance of 10 kpc).

As mentioned above, many of the neutron star SXTs show both thermal
and non-thermal components in their X-ray spectra, but the origin
of the non-thermal power-law component is uncertain.  In Figure~5
of \cite{jonker04}, it is shown that the spectra tend to be more
highly dominated by the power-law flux both below and above
$L\sim 10^{33}$ ergs~s$^{-1}$, but that the spectra measured 
when the luminosities are close to this range are dominated by
the thermal component.  The power-law on the higher luminosity
side is thought to be due to residual accretion, so the fact that
4U~1730--22 does not show a power-law may indicate that there is, 
at most, a low level of residual accretion.  Although it is unclear 
what causes the power-law on the lower luminosity side, it may be 
that all systems have weak power-law emission that is only detectable 
for the systems with the coolest neutron stars.

Given that 4U~1730--22 has not been seen in outburst since the 1970s, 
one could ask why the source would have one of the hotter (and thus 
more luminous) quiescent neutron stars.  Although the cooling time 
scale for the neutron star crust is on the order of years, the thermal 
time scale of the core is $\sim$10,000 years \citep{colpi01}, so a high 
quiescent temperature could simply indicate a higher level of activity 
over the last 10 millennia.  One hint that this might be the case is that 
the 1972 outburst from 4U~1730--22 lasted for $t_{\rm otb} \sim 230$ days, 
which is a factor of a few longer than is typically seen for neutron 
star SXTs \citep{csl97}.  Based on the theoretical work of \cite{bbr98}, 
the expression $t_{\rm rec} = (t_{\rm otb}/130)(F_{\rm otb}/F_{\rm q})$, 
can be used to estimate the outburst recurrence time ($t_{\rm rec}$) 
necessary to obtain the ratio of mean outburst flux ($F_{\rm otb}$) to 
quiescent flux ($F_{\rm q}$).  Using $t_{\rm otb} = 230$ days, 
$F_{\rm otb} = 10^{-9}$ ergs~cm$^{-2}$~s$^{-1}$ and the value of 
$F_{\rm q}$ given above, we estimate an recurrence time of $\sim$30 years.  
As outbursts from SXTs are not strictly periodic, this is consistent with 
the observed behavior to date, but it is also possible that the source had 
a higher level of activity in the past.

\section{Summary and Conclusions}

We have used {\em Chandra} and optical observations of the field of the 
{\em Uhuru} X-ray binary 4U~1730--22 to search for counterparts to this 
past X-ray transient.  Our analysis of the {\em Chandra} data indicates 
that we very likely have discovered the quiescent 4U~1730--22 X-ray 
counterpart.  The brightest X-ray source in or near the {\em Uhuru} error 
region (CXOU J173357.5--220156) has a purely thermal spectrum with an 
effective temperature of $131\pm 21$~eV, consistent with being a 10 km 
radius neutron star at a distance of $10^{+12}_{-4}$ kpc.  Such a 
temperature requires that the neutron star is very young ($<$10,000 years) 
or that it is maintaining its high temperature via accretion.  The lack of 
a supernova remnant suggests that the latter is the case, indicating that 
the system is a neutron star X-ray binary.  The likely association between 
CXOU J173357.5--220156 and the {\em Uhuru} source would indicate that 
4U~1730--22 harbors a neutron star.  The fact that no optical counterpart 
to the {\em Chandra} source is present to a limit of $R > 22.1$ is 
consistent with the source being an LMXB with K-type main sequence 
companion at a distance greater than a few kpc.  Deeper optical 
observations would be useful for learning more about this system.

The luminosity of the {\em Chandra} source that is the likely quiescent
counterpart to 4U~1730--22 of $L = 1.9\times 10^{33}$ ergs~s$^{-1}$ (assuming 
our estimated distance of 10 kpc) makes it one of the five most luminous of 
the 20 such systems for which measurements have been made.  This is interesting 
in the context of the comparison between neutron star and black hole luminosities 
as it is an example of a quiescent neutron star system that is significantly
more luminous than the known black hole systems.  In addition, this luminosity 
is in the range where \cite{jonker04} find that most neutron star SXTs have 
purely thermal spectra as we observe in this case.  Finally, a relatively
high level of X-ray outburst activity from 4U~1730--22 is required to maintain
the measured quiescent luminosity.  The system either needs to produce 
outbursts like its 1972 outburst every $\sim$30 years or it had a higher
level of activity at some point in the last $\sim$10,000 years.

\acknowledgments

JAT would like to thank E.~Armstrong and J.~Cooke for help with reducing
the optical data.  JAT acknowledges partial support from {\em Chandra} 
award number GO4-5052X issued by the {\em Chandra} X-ray Observatory 
Center, which is operated by the Smithsonian Astrophysical Observatory 
for and on behalf of NASA under contract NAS8-03060.  The SIMBAD database 
and the HEASARC Data Archive were used in preparing this paper.

% BIBLIOGRAPHY
%\bibliographystyle{jwapjbib}
%\bibliography{refs}

\begin{thebibliography}{}

\bibitem[\protect\astroncite{{Anders} \& {Grevesse}}{1989}]{ag89}
{Anders}, E., \& {Grevesse}, N.,  1989, Geochimica et Cosmochimica Acta, 53,
  197

\bibitem[\protect\astroncite{{Balucinska-Church} \& {McCammon}}{1992}]{bm92}
{Balucinska-Church}, M., \& {McCammon}, D.,  1992, ApJ, 400, 699

\bibitem[\protect\astroncite{{Brown}, {Bildsten} \& {Rutledge}}{1998}]{bbr98}
{Brown}, E.~F., {Bildsten}, L., \& {Rutledge}, R.~E.,  1998, ApJ, 504, L95

\bibitem[\protect\astroncite{{Campana} et~al.}{1998}]{campana98}
{Campana}, S., {Colpi}, M., {Mereghetti}, S., {Stella}, L., \& {Tavani}, M.,
  1998, A\&A~Rev., 8, 279

\bibitem[\protect\astroncite{{Canizares} et~al.}{2000}]{canizares00}
{Canizares}, C.~R., et~al., 2000, ApJ, 539, L41

\bibitem[\protect\astroncite{{Cardelli}, {Clayton} \& {Mathis}}{1989}]{ccm89}
{Cardelli}, J.~A., {Clayton}, G.~C., \& {Mathis}, J.~S.,  1989, \apj, 345, 245

\bibitem[\protect\astroncite{{Charles} \& {Coe}}{2006}]{cc06}
{Charles}, P.~A., \& {Coe}, M.~J.,  2006,
\newblock {Optical, ultraviolet and infrared observations of X-ray binaries},
\newblock  Compact stellar X-ray sources.~Edited by Walter Lewin \& Michiel van
  der Klis: Cambridge University Press),  215--265

\bibitem[\protect\astroncite{{Chen}, {Shrader} \& {Livio}}{1997}]{csl97}
{Chen}, W., {Shrader}, C.~R., \& {Livio}, M.,  1997, ApJ, 491, 312

\bibitem[\protect\astroncite{{Chevalier} et~al.}{1999}]{chevalier99}
{Chevalier}, C., {Ilovaisky}, S.~A., {Leisy}, P., \& {Patat}, F.,  1999, A\&A,
  347, L51

\bibitem[\protect\astroncite{{Colpi} et~al.}{2001}]{colpi01}
{Colpi}, M., {Geppert}, U., {Page}, D., \& {Possenti}, A.,  2001, ApJ, 548,
  L175

\bibitem[\protect\astroncite{{Cominsky} et~al.}{1978}]{cominsky78}
{Cominsky}, L., {Jones}, C., {Forman}, W., \& {Tananbaum}, H.,  1978, ApJ, 224,
  46

\bibitem[\protect\astroncite{{Corbel}, {Tomsick} \& {Kaaret}}{2006}]{ctk06}
{Corbel}, S., {Tomsick}, J.~A., \& {Kaaret}, P.,  2006, ApJ, 636, 971

\bibitem[\protect\astroncite{{Cox}}{2000}]{cox00}
{Cox}, A.~N.,  2000,
\newblock {Allen's astrophysical quantities},
\newblock  4th ed.~ Publisher: New York: AIP Press; Springer, 2000.~ Editedy by
  Arthur N.~Cox.)

\bibitem[\protect\astroncite{{Dickey} \& {Lockman}}{1990}]{dl90}
{Dickey}, J.~M., \& {Lockman}, F.~J.,  1990, ARA\&A, 28, 215

\bibitem[\protect\astroncite{{Forman} et~al.}{1978}]{forman78}
{Forman}, W., {Jones}, C., {Cominsky}, L., {Julien}, P., {Murray}, S.,
  {Peters}, G., {Tananbaum}, H., \& {Giacconi}, R.,  1978, ApJS, 38, 357

\bibitem[\protect\astroncite{{Freeman} et~al.}{2002}]{freeman02}
{Freeman}, P.~E., {Kashyap}, V., {Rosner}, R., \& {Lamb}, D.~Q.,  2002, ApJS,
  138, 185

\bibitem[\protect\astroncite{{Garcia} et~al.}{2001}]{garcia01}
{Garcia}, M.~R., {McClintock}, J.~E., {Narayan}, R., {Callanan}, P., {Barret},
  D., \& {Murray}, S.~S.,  2001, ApJ, 553, L47

\bibitem[\protect\astroncite{{Garmire} et~al.}{2003}]{garmire03}
{Garmire}, G.~P., {Bautz}, M.~W., {Ford}, P.~G., {Nousek}, J.~A., \& {Ricker},
  G.~R.,  2003,
\newblock in X-Ray and Gamma-Ray Telescopes and Instruments for Astronomy.
  Edited by Joachim E. Truemper, Harvey D. Tananbaum. Proceedings of the SPIE,
  Volume 4851, pp. 28-44 (2003)., 28

\bibitem[\protect\astroncite{{Giacconi} et~al.}{1971}]{giacconi71}
{Giacconi}, R., {Kellogg}, E., {Gorenstein}, P., {Gursky}, H., \& {Tananbaum},
  H.,  1971, ApJ, 165, L27

\bibitem[\protect\astroncite{{Jonker} et~al.}{2006}]{jonker06}
{Jonker}, P.~G., {Bassa}, C.~G., {Nelemans}, G., {Juett}, A.~M., {Brown},
  E.~F., \& {Chakrabarty}, D.,  2006, MNRAS, 368, 1803

\bibitem[\protect\astroncite{{Jonker} et~al.}{2004}]{jonker04}
{Jonker}, P.~G., {Galloway}, D.~K., {McClintock}, J.~E., {Buxton}, M.,
  {Garcia}, M., \& {Murray}, S.,  2004, MNRAS, 354, 666

\bibitem[\protect\astroncite{{Kong} et~al.}{2002}]{kong02a}
{Kong}, A.~K.~H., {McClintock}, J.~E., {Garcia}, M.~R., {Murray}, S.~S., \&
  {Barret}, D.,  2002, ApJ, 570, 277

\bibitem[\protect\astroncite{{Landolt}}{1992}]{landolt92}
{Landolt}, A.~U.,  1992, AJ, 104, 340

\bibitem[\protect\astroncite{{McClintock}, {Narayan} \&
  {Rybicki}}{2004}]{mnr04}
{McClintock}, J.~E., {Narayan}, R., \& {Rybicki}, G.~B.,  2004, ApJ, 615, 402

\bibitem[\protect\astroncite{{Narayan}, {Garcia} \& {McClintock}}{1997}]{ngm97}
{Narayan}, R., {Garcia}, M.~R., \& {McClintock}, J.~E.,  1997, ApJ, 478, L79

\bibitem[\protect\astroncite{{Pavlov}, {Shibanov} \& {Zavlin}}{1991}]{psz91}
{Pavlov}, G.~G., {Shibanov}, I.~A., \& {Zavlin}, V.~E.,  1991, MNRAS, 253, 193

\bibitem[\protect\astroncite{{Predehl} \& {Schmitt}}{1995}]{ps95}
{Predehl}, P., \& {Schmitt}, J. H. M.~M.,  1995, A\&A, 293, 889

\bibitem[\protect\astroncite{{Rutledge} et~al.}{1999}]{rutledge99}
{Rutledge}, R.~E., {Bildsten}, L., {Brown}, E.~F., {Pavlov}, G.~G., \&
  {Zavlin}, V.~E.,  1999, ApJ, 514, 945

\bibitem[\protect\astroncite{{Rutledge} et~al.}{2002}]{rutledge02}
{Rutledge}, R.~E., {Bildsten}, L., {Brown}, E.~F., {Pavlov}, G.~G., {Zavlin},
  V.~E., \& {Ushomirsky}, G.,  2002, ApJ, 580, 413

\bibitem[\protect\astroncite{{Tanaka} \& {Shibazaki}}{1996}]{ts96}
{Tanaka}, Y., \& {Shibazaki}, N.,  1996, ARA\&A, 34, 607

\bibitem[\protect\astroncite{{Tomsick} et~al.}{2004}]{tomsick04_2123}
{Tomsick}, J.~A., {Gelino}, D.~M., {Halpern}, J.~P., \& {Kaaret}, P.,  2004,
  ApJ, 610, 933

\bibitem[\protect\astroncite{{Tomsick}, {Gelino} \& {Kaaret}}{2005}]{tomsick05}
{Tomsick}, J.~A., {Gelino}, D.~M., \& {Kaaret}, P.,  2005, ApJ, 635, 1233

\bibitem[\protect\astroncite{{Tomsick} et~al.}{2002}]{tomsick02}
{Tomsick}, J.~A., {Heindl}, W.~A., {Chakrabarty}, D., \& {Kaaret}, P.,  2002,
  ApJ, 581, 570

\bibitem[\protect\astroncite{{Torres} et~al.}{2002}]{torres02}
{Torres}, M.~A.~P., {Casares}, J., {Mart{\' i}nez-Pais}, I.~G., \& {Charles},
  P.~A.,  2002, MNRAS, 334, 233

\bibitem[\protect\astroncite{{Tozzi} et~al.}{2006}]{tozzi06}
{Tozzi}, P., et~al., 2006, A\&A, 451, 457

\bibitem[\protect\astroncite{{Virani} et~al.}{2006}]{virani06}
{Virani}, S.~N., {Treister}, E., {Urry}, C.~M., \& {Gawiser}, E.,  2006, AJ,
  131, 2373

\bibitem[\protect\astroncite{{Weisskopf}}{2005}]{weisskopf05}
{Weisskopf}, M.~C.,  2005, astro-ph/0503091

\bibitem[\protect\astroncite{{Wijnands} et~al.}{2005}]{wijnands05}
{Wijnands}, R., {Homan}, J., {Heinke}, C.~O., {Miller}, J.~M., \& {Lewin},
  W.~H.~G.,  2005, ApJ, 619, 492

\bibitem[\protect\astroncite{{Wijnands} et~al.}{2004}]{wijnands04}
{Wijnands}, R., {Homan}, J., {Miller}, J.~M., \& {Lewin}, W.~H.~G.,  2004, ApJ,
  606, L61

\bibitem[\protect\astroncite{{Yakovlev} \& {Pethick}}{2004}]{yp04}
{Yakovlev}, D.~G., \& {Pethick}, C.~J.,  2004, ARA\&A, 42, 169

\bibitem[\protect\astroncite{{Zavlin}, {Pavlov} \& {Shibanov}}{1996}]{zps96}
{Zavlin}, V.~E., {Pavlov}, G.~G., \& {Shibanov}, Y.~A.,  1996, A\&A, 315, 141

\end{thebibliography}

% TABLES

\begin{table}
\caption{Detected {\em Chandra} Sources and Optical Identifications\label{tab:sources}}
\begin{minipage}{\linewidth}
\footnotesize
\begin{tabular}{lccccc} \hline \hline
Number & {\em Chandra} & {\em Chandra} & {\em Chandra} & X-ray/Optical & $R$-band\\
       & R.A. (J2000)  & Decl. (J2000) & Counts        & Separation    & Magnitude\\ \hline
 1\footnote{One of the 13 sources in the 90\% confidence {\em Uhuru} error region or within $30^{\prime\prime}$ of the error region.  Source 1 is CXOU J173357.5--220156, and source 2 is CXOU J173358.1--220101.}  & $17^{\rm h}33^{\rm m}57^{\rm s}\!.55$ & --$22^{\circ}01^{\prime}56^{\prime\prime}\!.9$ & 544.5 & --- & $>$22.1\\
 2$^{a}$ & $17^{\rm h}33^{\rm m}58^{\rm s}\!.18$ & --$22^{\circ}01^{\prime}01^{\prime\prime}\!.7$ &  87.8 & --- & $>$22.1\\
 3 & $17^{\rm h}34^{\rm m}11^{\rm s}\!.13$ & --$22^{\circ}03^{\prime}43^{\prime\prime}\!.8$ &  46.9 & --- & ---\\
 4 & $17^{\rm h}34^{\rm m}09^{\rm s}\!.13$ & --$22^{\circ}03^{\prime}33^{\prime\prime}\!.8$ &  32.6 & --- & ---\\
 5 & $17^{\rm h}33^{\rm m}48^{\rm s}\!.50$ & --$22^{\circ}00^{\prime}09^{\prime\prime}\!.0$ &  31.7 & --- & ---\\
 6 & $17^{\rm h}33^{\rm m}47^{\rm s}\!.33$ & --$22^{\circ}04^{\prime}03^{\prime\prime}\!.1$ &  30.8 & --- & ---\\
 7 & $17^{\rm h}34^{\rm m}00^{\rm s}\!.97$ & --$22^{\circ}00^{\prime}17^{\prime\prime}\!.2$ &  24.9 & --- & ---\\
 8 & $17^{\rm h}34^{\rm m}02^{\rm s}\!.81$ & --$22^{\circ}04^{\prime}49^{\prime\prime}\!.2$ &  19.4 & --- & ---\\
 9 & $17^{\rm h}34^{\rm m}14^{\rm s}\!.39$ & --$22^{\circ}01^{\prime}03^{\prime\prime}\!.0$ &  17.0 & --- & ---\\
10 & $17^{\rm h}34^{\rm m}11^{\rm s}\!.97$ & --$21^{\circ}59^{\prime}45^{\prime\prime}\!.6$ &  16.9 & --- & ---\\
11 & $17^{\rm h}33^{\rm m}46^{\rm s}\!.50$ & --$22^{\circ}03^{\prime}48^{\prime\prime}\!.6$ &  16.3 & --- & ---\\
12 & $17^{\rm h}34^{\rm m}02^{\rm s}\!.27$ & --$22^{\circ}00^{\prime}09^{\prime\prime}\!.7$ &  14.2 & --- & ---\\
13 & $17^{\rm h}34^{\rm m}05^{\rm s}\!.97$ & --$22^{\circ}00^{\prime}07^{\prime\prime}\!.8$ &  13.1 & --- & ---\\
14$^{a}$ & $17^{\rm h}34^{\rm m}06^{\rm s}\!.99$ & --$22^{\circ}02^{\prime}14^{\prime\prime}\!.1$ &  12.1 & --- & $>$22.1\\
15$^{a}$ & $17^{\rm h}34^{\rm m}01^{\rm s}\!.31$ & --$22^{\circ}01^{\prime}45^{\prime\prime}\!.4$ &  11.1 & --- & $>$22.1\\
16 & $17^{\rm h}33^{\rm m}59^{\rm s}\!.47$ & --$21^{\circ}59^{\prime}13^{\prime\prime}\!.1$ &  11.1 & --- & ---\\
17 & $17^{\rm h}33^{\rm m}59^{\rm s}\!.55$ & --$21^{\circ}59^{\prime}29^{\prime\prime}\!.3$ &  11.0 & --- & ---\\
18$^{a}$ & $17^{\rm h}33^{\rm m}55^{\rm s}\!.91$ & --$22^{\circ}01^{\prime}37^{\prime\prime}\!.6$ &  10.1 & $0^{\prime\prime}\!.58$ & $16.7\pm 0.1$\\
19$^{a}$ & $17^{\rm h}34^{\rm m}02^{\rm s}\!.54$ & --$22^{\circ}02^{\prime}42^{\prime\prime}\!.5$ &  10.0 & $0^{\prime\prime}\!.20$ & $18.9\pm 0.1$\\
20 & $17^{\rm h}34^{\rm m}04^{\rm s}\!.61$ & --$22^{\circ}00^{\prime}48^{\prime\prime}\!.6$ &   9.6 & --- & ---\\
21 & $17^{\rm h}34^{\rm m}10^{\rm s}\!.84$ & --$22^{\circ}00^{\prime}34^{\prime\prime}\!.9$ &   9.0 & --- & ---\\
22 & $17^{\rm h}34^{\rm m}07^{\rm s}\!.49$ & --$22^{\circ}00^{\prime}33^{\prime\prime}\!.0$ &   8.7 & --- & ---\\
23 & $17^{\rm h}34^{\rm m}09^{\rm s}\!.49$ & --$22^{\circ}04^{\prime}31^{\prime\prime}\!.5$ &   8.4 & --- & ---\\
24$^{a}$ & $17^{\rm h}33^{\rm m}49^{\rm s}\!.66$ & --$22^{\circ}02^{\prime}21^{\prime\prime}\!.9$ &   8.2 & --- & $>$22.1\\
25$^{a}$ & $17^{\rm h}34^{\rm m}03^{\rm s}\!.96$ & --$22^{\circ}03^{\prime}40^{\prime\prime}\!.0$ &   7.6 & --- & $>$22.1\\
26$^{a}$ & $17^{\rm h}33^{\rm m}56^{\rm s}\!.74$ & --$22^{\circ}00^{\prime}45^{\prime\prime}\!.6$ &   7.4 & $0^{\prime\prime}\!.50$ & $21.5\pm 0.5$\\
27 & $17^{\rm h}34^{\rm m}01^{\rm s}\!.68$ & --$22^{\circ}00^{\prime}40^{\prime\prime}\!.9$ &   7.0 & --- & ---\\
28$^{a}$ & $17^{\rm h}33^{\rm m}50^{\rm s}\!.53$ & --$22^{\circ}01^{\prime}45^{\prime\prime}\!.9$ &   6.4 & --- & $>$22.1\\
29 & $17^{\rm h}34^{\rm m}14^{\rm s}\!.85$ & --$22^{\circ}02^{\prime}58^{\prime\prime}\!.2$ &   6.2 & --- & ---\\
30 & $17^{\rm h}34^{\rm m}04^{\rm s}\!.84$ & --$21^{\circ}59^{\prime}57^{\prime\prime}\!.6$ &   5.9 & --- & ---\\
31$^{a}$ & $17^{\rm h}33^{\rm m}52^{\rm s}\!.88$ & --$22^{\circ}02^{\prime}17^{\prime\prime}\!.3$ &   5.9 & $0^{\prime\prime}\!.49$ & $21.5\pm 0.5$\\
32 & $17^{\rm h}34^{\rm m}03^{\rm s}\!.10$ & --$22^{\circ}00^{\prime}29^{\prime\prime}\!.8$ &   5.4 & --- & ---\\
33$^{a}$ & $17^{\rm h}34^{\rm m}05^{\rm s}\!.91$ & --$22^{\circ}01^{\prime}43^{\prime\prime}\!.9$ &   4.5 & $0^{\prime\prime}\!.55$ & $21.5\pm 0.5$\\
34 & $17^{\rm h}34^{\rm m}05^{\rm s}\!.47$ & --$21^{\circ}59^{\prime}36^{\prime\prime}\!.8$ &   4.5 & --- & ---\\
35$^{a}$ & $17^{\rm h}34^{\rm m}03^{\rm s}\!.15$ & --$22^{\circ}01^{\prime}55^{\prime\prime}\!.5$ &   4.4 & --- & $>$22.1\\ \hline
\tablecomments{For all sources, the largest contribution to the uncertainties in the {\em Chandra} positions are due to pointing systematics.  The pointing uncertainties are $0^{\prime\prime}\!.64$ at 90\% confidence and $1^{\prime\prime}$ at 99\% confidence \citep{weisskopf05}.  The statistical position uncertainties depend on source brightness and off-axis angle.  For most sources, the statistical uncertainties are between $0^{\prime\prime}\!.1$ and $0^{\prime\prime}\!.2$, and they are $<$$0^{\prime\prime}\!.4$ for all 35 sources.}
\end{tabular}
\end{minipage}
\end{table}

\begin{table}
\caption{Spectral Fits\label{tab:spectra}}
\begin{minipage}{\linewidth}
\footnotesize
\begin{tabular}{lcccccc} \hline \hline
Model & $N_{\rm H}$\footnote{The column density in units of $10^{22}$ cm$^{-2}$.} & 
$\Gamma$ & $F_{\rm pl}$\footnote{Unabsorbed 0.3--8 keV power-law flux in units of 
$10^{-12}$ ergs~cm$^{-2}$~s$^{-1}$.} & $\log{T_{\rm eff}}$\footnote{The base-10 logarithm of 
the neutron star's effective temperature in K.} & $N_{\rm NSA}$\footnote{The normalization
for the NSA model, which is equal to $d^{-2}$ where $d$ is the distance to the source in pc.}
& $\chi^{2}/\nu$ \\ \hline
\multicolumn{7}{c}{Source 1}\\ \hline
pl      & $0.86^{+0.14}_{-0.12}$ & $5.4^{+0.6}_{-0.5}$ & $7^{+11}_{-4}$ & --- & --- & 11.5/15\\
NSA     & $0.37^{+0.07}_{-0.12}$ & --- & --- & $6.18\pm 0.07$ & $(1.0^{+1.7}_{-0.8})\times 10^{-8}$ & 11.5/15\\
pl+NSA  & --- & --- & --- & --- & --- & 10.5/13\\ \hline
\multicolumn{7}{c}{Source 2}\\ \hline
pl      & $0.7^{+0.5}_{-0.3}$ & $2.1^{+1.2}_{-0.9}$ & $4^{+10}_{-1}$ & --- & --- & 12.4/5\\
NSA     & $0.3^{+0.4}_{-0.2}$ & --- & --- & 6.7--7.0\footnote{A range is given here because the model is not valid for values of $\log{T_{\rm eff}}$ above 7.0.} & (4.5--300)$\times 10^{-13}$ & 16.7/5\\
pl+NSA  & $2.2^{+2.1}_{-1.5}$ & $1.6^{+1.0}_{-1.2}$ & $0.038^{+0.158}_{-0.014}$ & $5.6^{+0.7}_{-0.2}$ & $(1,128,448^{+26,000,000}_{-1,128,444})\times 10^{-10}$ & 4.7/3\\ \hline
\end{tabular}
\end{minipage}
\tablecomments{The models used are a power-law (pl) and the Neutron Star Atmosphere (NSA) 
model of Pavlov et al. (1991)\nocite{psz91} and Zavlin et al. (1996)\nocite{zps96}.  
For the NSA model, we have fixed the neutron star radius to 10 km and the neutron star 
mass to 1.4\Msun.  Also, a low magnetic field ($B < 10^{8} - 10^{9}$ G) is assumed.
Quoted errors are 90\% confidence for all parameters.}
\end{table}

\end{document}